# Bias in the Journal Impact Factor

*Jerome K Vanclay*
*Southern Cross University, PO Box 157, Lismore NSW 2480, Australia*
*Tel +61 2 6620 3147, Fax +61 2 6621 2668, JVanclay@scu.edu.au*

**Abstract**

The ISI journal impact factor (JIF) is based on a sample that may represent half the whole-of-life citations to some journals, but a small fraction (<10%) of the citations accruing to other journals. This disproportionate sampling means that the JIF provides a misleading indication of the true impact of journals, biased in favour of journals that have a rapid rather than a prolonged impact. Many journals exhibit a consistent pattern of citation accrual from year to year, so it may be possible to adjust the JIF to provide a more reliable indication of a journal's impact.

**Introduction**

Despite well-recognised limitations (e.g., Dong et al. 2005, Kaltenborn and Kuhn 2004, Leeuwen et al. 1999, Moed 1999, Seglen 1997), the Institute for Scientific Information's (ISI) journal impact factor (JIF) continues to influence scientific endeavour (Jennings 1998, Weingart 2005). This disadvantages some disciplines, because bias is implicit in the citation sample on which the JIF is based. This bias has previously been noticed by Moed et al. (1998, 1999). The JIF is based on the number of citations accruing during a given year ($i$), to journal issues published in the two preceding years ($i-1$ and $i-2$). Thus, a journal contribution has a two-year window, namely the first and second years after publication, during which it may contribute towards the journal's impact factor. This two-year window may sample a large proportion of citations to some contributions, but a small sample of citations to other material. This paper illustrates the extent of this bias and suggests one possible remedy.

Glänzel and Schoepflin (1995) and Moed et al. (1998) have shown that the nature of this bias depends on both the journal and the field of endeavour. Several researchers (e.g., Egghe and Rousseau 2000) have examined patterns of citations to individual papers. The present study is complementary to these, as it is not concerned with citations to individual papers, but with citations to all the material published by a journal in a given year. The JIF does not deal evenly with 'Hares' (journals to which citations accrue quickly over a confined period) and 'Tortoises' (journals to which citations accrue slowly over an extended period), because the 2-year sample represents a much larger proportion of total citations for the Hares. For some Hares illustrated below, the JIF's two-year window may sample half the total citations, whereas the same window may sample fewer than one tenth of the total citations to Tortoises. Unlike the fable in which the slow and steady tortoise wins the race, the JIF offers the accolades to the hare. The magnitude of this bias is illustrated with data on two journals drawn from ISI's Web of Science (WoS). I am concerned only with the

JIF numerator, the number of citations received by a journal, because others (e.g., Jacso 2001) have previously offered suggestions to improve the denominator (the number of citeable items).

**Journal Selection**

I created a list of journal titles that were indexed by ISI continuously since 1992, and accrued fewer than 2000 citations during 2002-3, the observation period for the 2004 Journal Citation Report (JCR). The limit of 2000 citations was imposed because of limitations in the analytical capacity offered through WoS. To ensure broad coverage of a wide range of citation patterns, I stratified this list according to discipline, JIF, immediacy index, total citations and cited half-life using data from JCR 2004. I selected 14 titles (*Acm T Math Software, Adv Agron, Ann Appl Biol, Adv Phys, Atom Energy, Ca Cancer J Clin, J Plant Growth Regul, J Wildlife Dis, Mass Spectrom Rev, Neurosurg Clin N Am, Newsl Stratigr, North J Appl For, Opt Mater, Scientist*) from this list, and plotted the annual accrual of citations to items published in 1992 (Figure 1). From these 14 titles, I subjectively selected two journals (*Acm T Math Software*, here denoted the Tortoise, and *Scientist*, denoted the Hare) with a comparable number of citations in 1994, but with contrasting trends of citation accrual. According to the 2004 JCR, these two journals, denoted the Hare and the Tortoise, had JIFs of 0.2 and 1.3, and cited half-lives of 2.4 and >10 respectively. While the Hare is noteworthy for the abrupt culmination of the citation accrual pattern, it is evident from Figure 1 that the selected Hare and Tortoise are not unique, and that there are several other journal combinations that are comparable in terms of cumulative citations in 1993 or 1994, but very different citation totals by 2005.

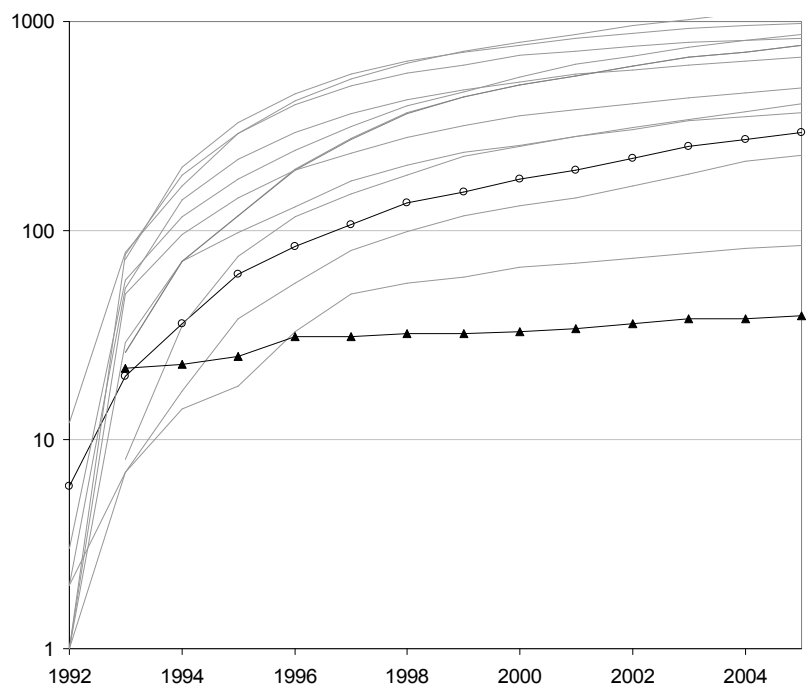

**Figure 1**. The citation accrual patterns to 14 journals selected from the WoS. The Hare (▲) and the Tortoise (○) are highlighted.

**Citations to two Selected Journals**

The graph of cumulative citations 1987-2005 for the Tortoise reveals a series of near-parallel lines, showing a steady accumulation of citations to each year of publication, with little evidence of declining interest in earlier papers, or of citation inflation (Figure 2). In Figure 2, citations accruing to the 1996 volume of the Tortoise are more numerous (dotted line), because of one particularly noteworthy paper, but this anomaly is resolved by standardizing to the citation count in the second year after publication (the year that usually contributes most to the JIF; so that cumulative citations accrued by the $2^{nd}$ year correspond to 100%; Figure 3). Figure 3 illustrates that, for the Tortoise, counting citations over the conventional two-year window (years *i-1* and *i-2*) gives a good indication of long-term trends, and shows that this trend continues unabated for many years.

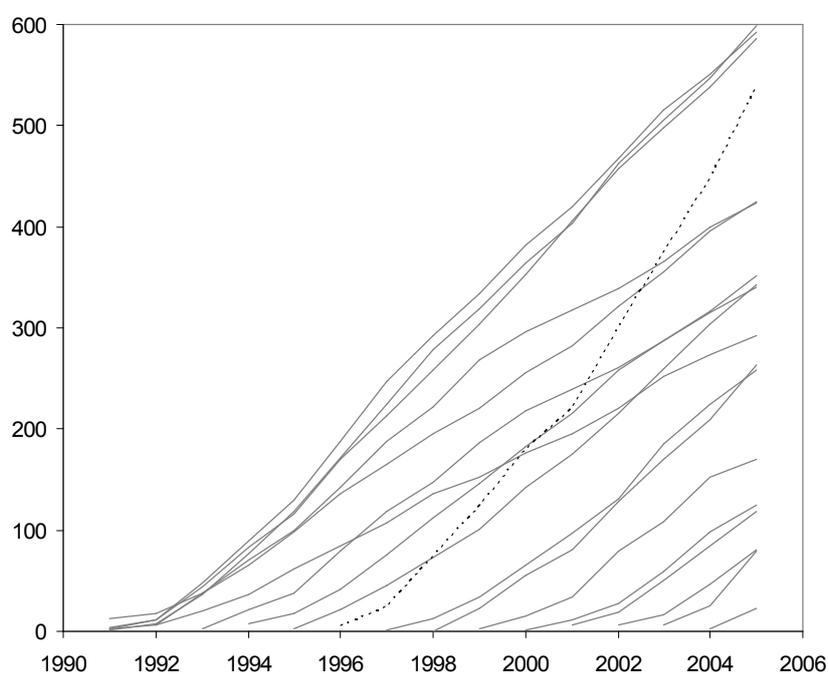

**Figure 2**. Citations accruing to each year of publication of the Tortoise. Trends are relatively consistent, apart from the 1996 volume (dotted line) that carried a particularly noteworthy paper with 162 citations. Data from WoS.

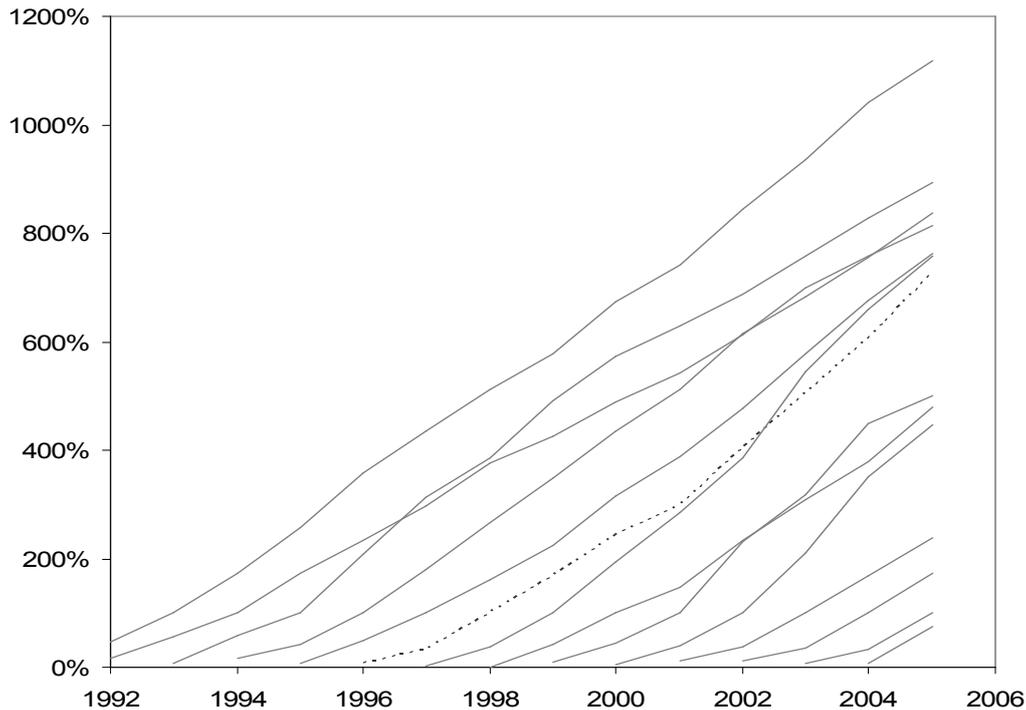

**Figure 3**. When cumulative citation counts to the Tortoise are standardized to year 2 (100%), a relatively consistent pattern of citation accrual emerges.

A similar trend emerges for the Hare, with a series of near-parallel lines during 1990-97, but with two conspicuous anomalies: citations accruing in 1994 to the 1993 volume of the Hare seem inflated, and citation accumulation trends become steeper after 1997 (Figure 4). It is not clear why this pattern emerges, but it may be that a citation perturbation may have contributed to an increased JIF in 1994 (published in 1995), stimulated additional contributions, and in turn, increased citations from 1997. Evidence supporting this view draws on the observation that 21 of 32 citations contributing to the journal's 1993 JIF were self-references (i.e., papers referring to other papers in the same journal). However, Lange (2002) investigated the impact of an erroneous JIF in another journal, and concluded that such impacts were likely to be small.

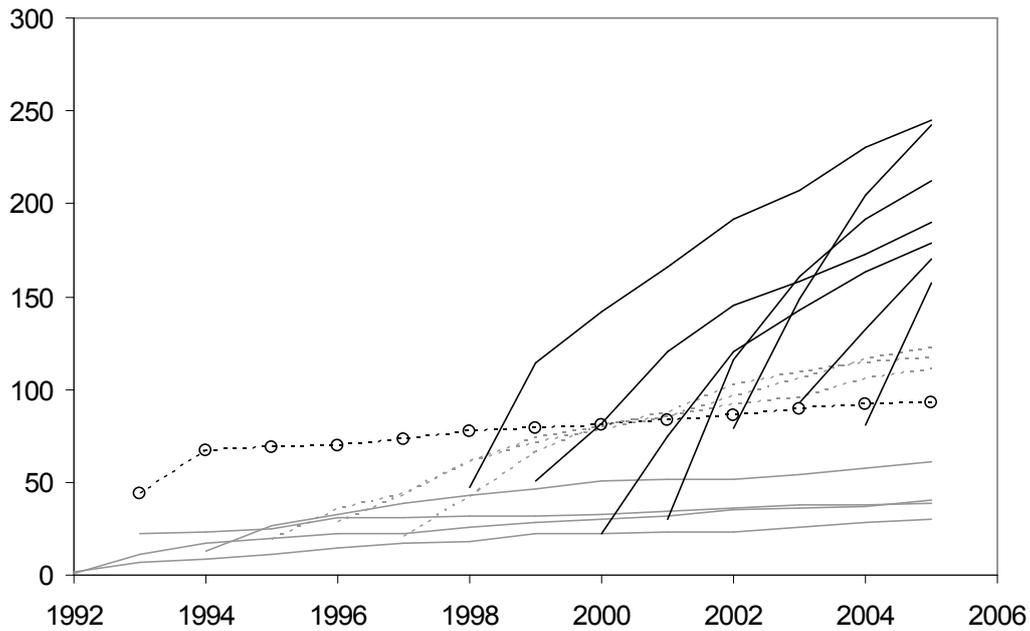

**Figure 4**. The Hare does not show the same consistent pattern of citation accrual. Citations to the 1993 volume (dotted, ○) are anomalous. Three different gradients are evident: 1990-1994 (grey, little slope), 1995-97 (dotted, intermediate slope), and 1998-2004 (black, steep slope).

Most of these anomalies in the citation accrual patterns to the Hare can be removed by indexing the data to the second year after publication, but 1993 still stands out as anomalous (Figure 5). This may be attributed to the high rate of self-referencing: in 1993, 38 out of 44 citations were self-references to the 1993 issue. This high rate of self-referencing is not a record (The *World Journal of Gastroenterology* was suspended by ISI when self-referencing reached 85%, Monastersky 2005), but it appears to have had a substantial effect on the future success of the journal. Removing these self-references leads to more consistent trends (Figure 6).

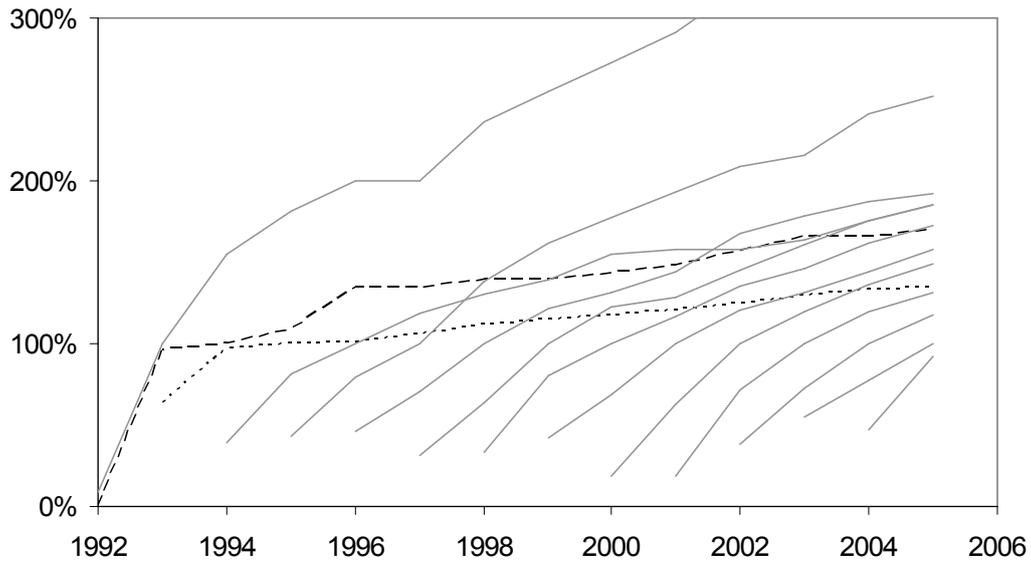

**Figure 5**. Standardizing to year 2 creates more consistent trends, apart from the 1992 (dashed) and 1993 (dotted) issues with many self-references.

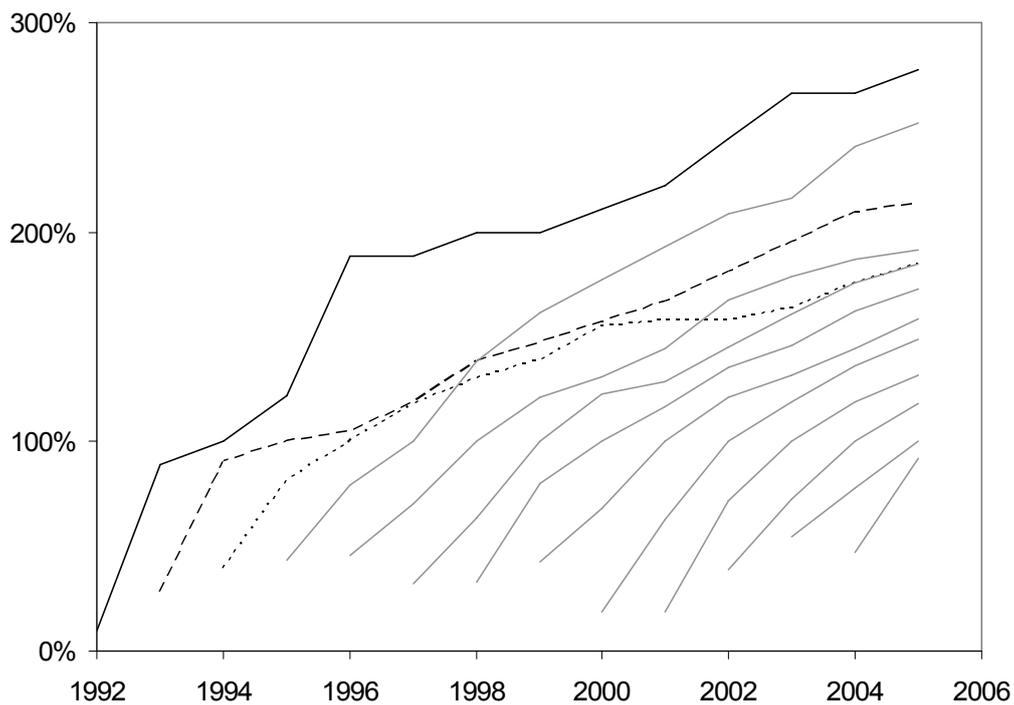

**Figure 6**. Removing self-references leads to a more consistent pattern of citation accrual to the Hare.

Despite minor anomalies, the standardised cumulative citation curves seem to be relatively consistent from year to year, and characteristic to each journal (Figure 7). Figure 7 also exposes the magnitude of the bias in the JIF, illustrating that citations accruing in years 1 and 2, constitute about 40% of the lifetime citations to the Hare, but less than 10% of the citations accruing to the Tortoise.

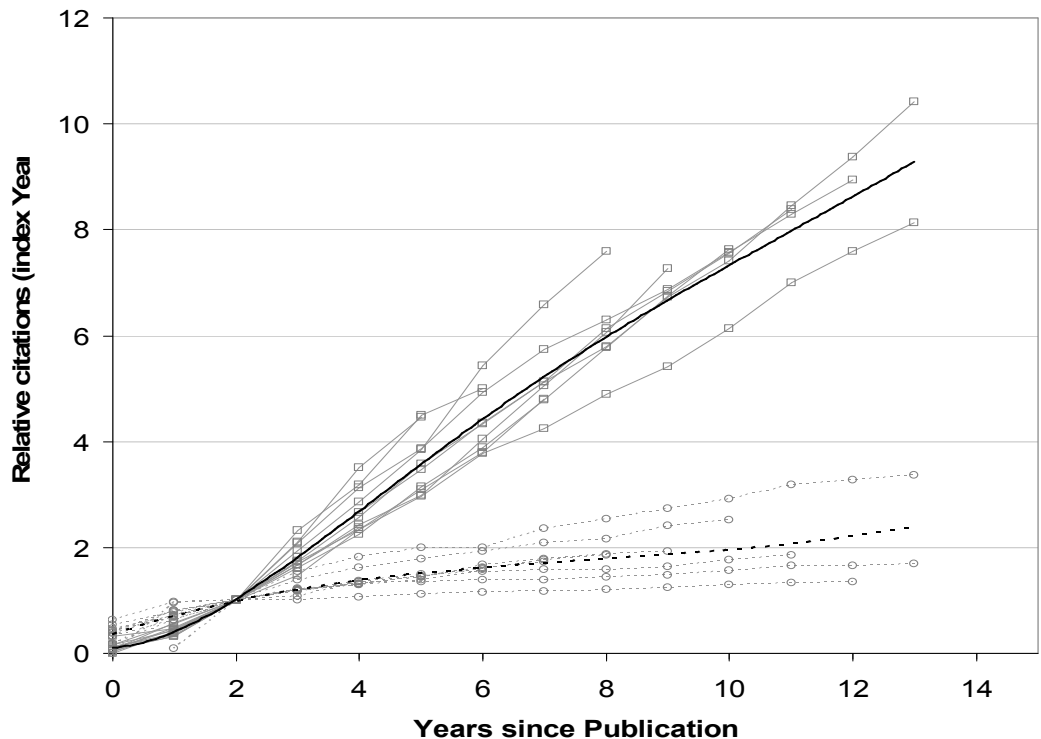

**Figure 7**. Standardized data from the Hare and the Tortise, illustrating the differences between the gradients of each.

**Adjusting the JIF to reduce bias**

The 2-year window used by ISI to gauge JIF creates a large distortion in favour of the Hare. Because the JIF appears entrenched (Monastersky 2005), it should be standardised to better represent the relative impact of Hares and Tortoises. A 3-year (Glänzel and Schoepflin 1995) or larger window (Moed et al. 1998) will not solve this bias, and a variable window based on the cited half-life (Sombatsompop et al. 2004) will create new difficulties with time-lags that interfere with timely comparisons between journals. However, it is possible to scale the 2-year JIF to better approximate a standardised proportion of citations (Figure 8), or to normalise the JIF within fields of endeavour (Moed et al. 1998, 1999).

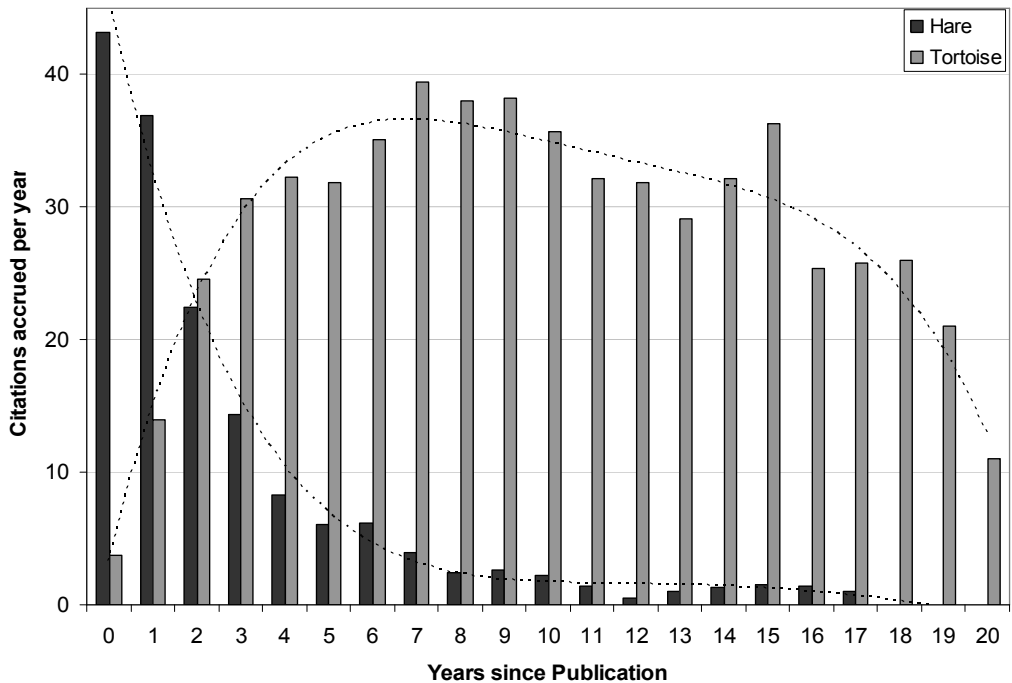

**Figure 8**. Average number of citations accruing each year to two journals denoted the Hare and the Tortoise. Only years 1 & 2 contribute to the JIF; these represent 38% of the 20-year total in the case of the Hare, but only 6% in the case of the Tortoise.

Figure 8 illustrates the average number of citations accruing to each journal each year after publication. It needs to be interpreted with some caution, because the sample size varies inversely with year after publication (i.e., the year 0 datum is the mean of 14 observations of citations accruing to volumes published during 1994-2005, whereas the year 20 datum represents a single instance of citations in 2006 to the 1986 volume), but it does give a good insight into the nature of the bias. The 2-year window sampled by the JIF represents 38% of 'life-time' (actually 20-year) citations to the Hare (comparable to several other journals identified by Decker and Brähler 2001), but only about 6% of life-time citations to the Tortoise.

To scale the conventional JIF to represent the half-life (Sombatsompop et al. 2004) the published JIFs should be multiplied by 1.3 for the Hare, and by 7.7 for the Tortoise. This would change the estimated impact from 0.2 and 1.3 (the standard 2-year JIF) to 0.3 and 10.1 respectively for the estimated half-life impact. These scaling factors represent extremes, because this study has deliberately selected journals with contrasting patters of citation accrual, but it nonetheless serves to illustrate how the JIF is biased in favour of journals like the Hare. Whether the JIF is scaled to represent the half-life, or to any other decile is immaterial, but it is imperative that such an adjustment is made to level the playing field and create a fair comparison between research by molecular biologists and field ecologists, and by researchers of short-lived (e.g., fruit flies, *Drosophila*) and long-lived organisms (e.g., elephants, *Loxodonta*).

Such scaling is important if valid comparisons are to be made across a range of disciplines. It could be achieved by normalizing JIFs within fields of endeavour (Moed et al. 1999), or by scaling the JIF to represent equal sample sizes (as illustrated above). ISI could facilitate such adjustments by providing better estimates of the

'cited half-life' by (1) reporting the actual half-life (instead of truncating the estimate at 10 years) (Moed et al 1998), (2) by flagging instances where the journal age is less than twice the apparent half-life; and (3) taking care to account correctly for name-changes and mergers of journals (Leeuwen et al. 1999).

**Conclusion**

The two-year window used to estimate the JIF creates an unequal sample for different journals, and introduces a bias that means the JIF does not provide a comparable indication of impact across different disciplines. If such a comparison is required, the JIF should be adjusted to account for this unequal sample size.

---

**Notes**:
The Tortoise is *ACM Transactions on Mathematical Software* (ISSN 0098-3500).
The Hare is *The Scientist* (ISSN 0890-3670).